\documentclass[twocolumn,aps,prc,superscriptaddress,nofootinbib,showpacs,floatfix]{revtex4-1}
\usepackage{url}
\usepackage{cancel}
\usepackage[colorlinks,linkcolor=blue,citecolor=blue,filecolor=black,urlcolor=blue]{hyperref}
\usepackage{epsfig,graphics}
\usepackage{graphicx}% Include figure files
\usepackage{dcolumn}% Align table columns on decimal point
\usepackage{bm}% bold math
\usepackage[usenames]{color}
\usepackage{amssymb}
\usepackage{amsmath}
\usepackage{multirow}
\usepackage{float}
\usepackage{harpoon}
\usepackage{MnSymbol}
\usepackage{appendix}
\usepackage{color}
\makeatletter

\newcommand{\Rmnum}[1]{\expandafter\@slowromancap\romannumeral #1@}
\makeatother

\makeatletter % `@' now normal "letter"
\@addtoreset{equation}{section}
\makeatother  % `@' is restored as "non-letter"

\newcommand{\sqrtsnn}{\mbox{$\sqrt{s_{\mathrm{NN}}}$}}
\newcommand{\npart}{\mbox{$N_{\mathrm{part}}$}}
\newcommand{\ncoll}{\mbox{$N_{\mathrm{coll}}$}}
\newcommand{\pT} {p_{\mathrm{T}}}

\newcommand{\lr}[1]{\left\langle #1\right\rangle}

\newcommand{\Dphi}{\mbox{$\Delta \phi$}}
\newcommand{\Deta}{\mbox{$\Delta \eta$}}
\newcommand{\nch}{\lr{N_{\mathrm{ch}}}}
\newcommand{\nchb}{N_{\mathrm{ch}}}

\usepackage{CJK}
\hfuzz=\maxdimen
\begin{document}
\title{Disentangle contributions to small-system collectivity via scans of light nucleus-nucleus collisions}
\newcommand{\sbu}{Department of Chemistry, Stony Brook University, Stony Brook, NY 11794, USA}
\newcommand{\bnl}{Physics Department, Brookhaven National Laboratory, Upton, NY 11976, USA}
\newcommand{\ccnu}{Institute of Particle Physics, Central China Normal University, Wuhan 430079, China}
\newcommand{\rice}{Physics Department, Rice University, Houston, Texas 77251, USA}
 \author{Shengli Huang}\affiliation{\sbu} \author{Zhenyu Chen}\affiliation{\sbu}\affiliation{\bnl}\author{Wei Li}\affiliation{\rice}\author{Jiangyong Jia}\affiliation{\sbu}\affiliation{\bnl}
\begin{abstract}
The observation of multi-particle azimuthal correlations in high-energy small-system collisions has led to intense debate on its origin and the possible coexistence from two competing theoretical scenarios: one based on initial-state intrinsic momentum anisotropy (ISM), and the other based on final-state collective response to the collision geometry (FSM). To complement the previous scan of asymmetric collision systems ($p$+Au, $d$+Au and He+Au), we propose a scan of small symmetric collision systems at RHIC, such as C+C, O+O, Al+Al and Ar+Ar $\sqrt{s_{\mathrm{NN}}}=0.2$ TeV, to provide further insights in disentangling contributions from these two scenarios. These symmetric small systems have the advantage of providing a better controlled initial geometry dominated by the average shape of the overlap region, as opposed to fluctuation-driven geometries in asymmetric systems. A transport model is employed to investigate the expected geometry response in the FSM scenario. Different trends of elliptic flow with increasing charge particle multiplicity are observed between symmetric and asymmetric systems, while triangular flow appears to show a similar behavior. Furthermore, a comparison of O+O collisions at $\sqrt{s_{\mathrm{NN}}}=0.2$  TeV and at $\sqrt{s_{\mathrm{NN}}}=2.76-7$ TeV, as proposed at the LHC, provides a unique opportunity to disentangle the collision geometry effects at nucleon level from those arising from subnucleon fluctuations.
\end{abstract}

%\pacs{25.75.Gz, 25.75.Ld, 25.75.-1}
%	25.75.Gz, Particle correlations and fluctuations
%	25.75.Ld,	Collective flow, relativistic collisions.
%      25.75.-q,	Relativistic heavy-ion collisions                
\maketitle
In high-energy proton-proton ($pp$), proton-nucleus ($p$+A) and nucleus-nucleus (A+A) collisions, particle correlations are important tools to study multi-parton dynamics of quantum chromodynamics (QCD) in the strongly-coupled, non-perturbative regime~\cite{Shuryak:2014zxa}. Measurements of azimuthal correlations reveal a strong harmonic modulation of particle densities d$N/{\textrm d}\phi\propto 1+2\sum_{n=1}^{\infty}v_{n}\cos n(\phi-\Phi_{n})$~\cite{Ollitrault:1992bk,Jia:2014jca,Dusling:2015gta}, where $v_n$ and $\Phi_n$ represent the magnitude and phase of the $n^{\mathrm{th}}$-order harmonic, and are often denoted by flow vector $V_n=v_n{\mathrm e}^{{\textrm i}n\Phi_n}$. The azimuthal correlations are found to be collective, involving many particles over a wide pseudorapidity range. The collectivity in A+A collisions is successfully described as a hydrodynamic response of the produced system to shape fluctuations in the initial state~\cite{Heinz:2013th}. However, such interpretation is challenged in small-system collisions such as $pp$ and $p$+A, where the small size and short lifetime might prevent the system to thermalize and evolve hydrodynamically. Instead, collectivity arising either from initial momentum correlation~\cite{Dusling:2015gta} or via a few scatterings among partons (without hydrodynamization)~\cite{He:2015hfa,Kurkela:2018qeb,Romatschke:2018wgi} has been proposed as alternative source of collectivity in small systems. Lots of experimental and theoretical efforts have been devoted to the study of collectivity in small-system collisions, with the goal of understanding the time-scale and origin for the emergence of collectivity and the mechanism for early-time thermalization in large collision systems. 

One key feature that distinguishes initial momentum correlation models (ISM) from final-state interaction models (FSM, including hydrodynamics or a few scatterings, denoted as FSM-hydro or FSM-tran.) is the relation between the initial-state geometry and final-state collectivity~\cite{Nagle:2018nvi}. In FSM, the collectivity is a geometrical response to initial shape fluctuations, i.e. $v_n$ is approximately proportional to the $n^{\textrm{th}}$-order initial-state eccentricity $\varepsilon_n$~\cite{Bozek:2011if}. In ISM, such geometrical response is expected to be absent~\cite{Schenke:2015aqa}. One idea to distinguish these two scenarios is to perform a geometry scan by colliding systems with different spacial eccentricities and see if the measured $v_n$ is correlated with the change of $\varepsilon_n$ between different systems~\cite{Nagle:2013lja}.

Several studies of elliptic flow ($v_2$) and triangular flow ($v_3$) based on such geometry scan have been performed at RHIC with $p$+Au, $d$+Au and $^3$He+Au~\cite{Adare:2013piz,PHENIX:2018lia,Huang:2019rsn,Lim:2018huo,Bozek:2014cya}. In high-multiplicity events, the $\varepsilon_2$ was predicted to be larger than in $p$+Au while the $\varepsilon_3$ is comparable~\cite{Nagle:2013lja}. Therefore, a similar hierarchy is expected for $v_2$ and $v_3$ in FSM, as observed experimentally~\cite{PHENIX:2018lia}. However, ISM based on a particular implementation of gluon saturation physics could produce large momentum anisotropy in these systems~\cite{Schenke:2015aqa}. The situation is more challenging in the understanding of collectivity involving heavy quarks, such as $D$ meson or $J/\Psi$ in $p$+Pb collisions~\cite{Sirunyan:2018kiz,Acharya:2017tfn,Sirunyan:2018toe}: FSM presently significantly underestimates the $v_2$ for $D$ and $J/\Psi$ ~\cite{Du:2018wsj}, while an ISM-based approach is able to describe the data~\cite{Zhang:2019dth}. The relative contribution of FSM vs. ISM for the $v_n$ data in small systems is an area of intense ongoing debate~\cite{smallsys,Loizides:2016tew}. Even in FSM, there are uncertainties in modeling the initial-state geometry due to different treatments of subnucleonic fluctuations, which are expected to play an important role especially in small asymmetric systems.  Furturemore, experimental studies from previous $p$/$d$/$^3$He+Au scan at RHIC were limited by detector capabilities: 1) most measurements were based on two-particle correlations with incomplete understanding of non-flow systematics (e.g., results depend strongly on the non-flow estimation method), 2) the nature of longitudinal decorrelations of collectivity and its effects on the measurements were poorly understood, 3) a large class of multi-particle observables, demonstrated to be very insightful at the LHC~\cite{Citron:2018lsq}, were only partially explored (multi-particle $v_{2}$ has been measured in d+Au~\cite{Aidala:2017ajz} but without $\pT$ differential information).

In this paper, an extended scan of small-system collisions at RHIC is proposed, taking advantage of the newly completed detector upgrades at STAR and future detector capabilities at STAR/sPHENIX (A case at LHC is studied in Ref.~\cite{Sievert:2019zjr} for a few collision species). We note that RHIC and LHC have collided many systems, but there is a large gap between the largest small system $^3$He+Au and smallest large system Cu+Cu. We propose additional system scans to fill the gap between $pp$ and Cu+Cu, in particular symmetric collision systems such as C+C, O+O, Al+Al, Ar+Ar. Since the system created in Cu+Cu collisions clearly exhibits final-state effects associated with the quark-gluon plasma (QGP), such as collective flow~\cite{Abelev:2010tr,Adare:2014bga} and jet quenching~\cite{Adare:2008ad,Abelev:2009ab}, a scan of smaller symmetric systems could help to establish at which system size initial-state effects become subdominant compared to final-state effects, as well as provide important level-arm to disentangle between the two final-state scenarios: FSM-hydro vs FSM-tran. Furthermore, the role of $\varepsilon_2$ and $\varepsilon_3$ in small A+A systems are very different from those in $p$/$d$/$^3$He+Au: the $\varepsilon_2$ has a significant average geometry component in small A+A systems, while it is dominated by fluctuations in $p$/$d$/$^3$He+Au systems. Therefore, different centrality dependence of $v_2$ is expected for symmetric and asymmetric systems. As argued in Ref.~\cite{Citron:2018lsq}, symmetric systems also have better centrality resolution and therefore less centrality bias compared to asymmetric systems, thanks to a broader distribution in the number of participating nucleons, $\npart$.

\begin{figure}[h!]
\begin{center}
\includegraphics[width=1\linewidth]{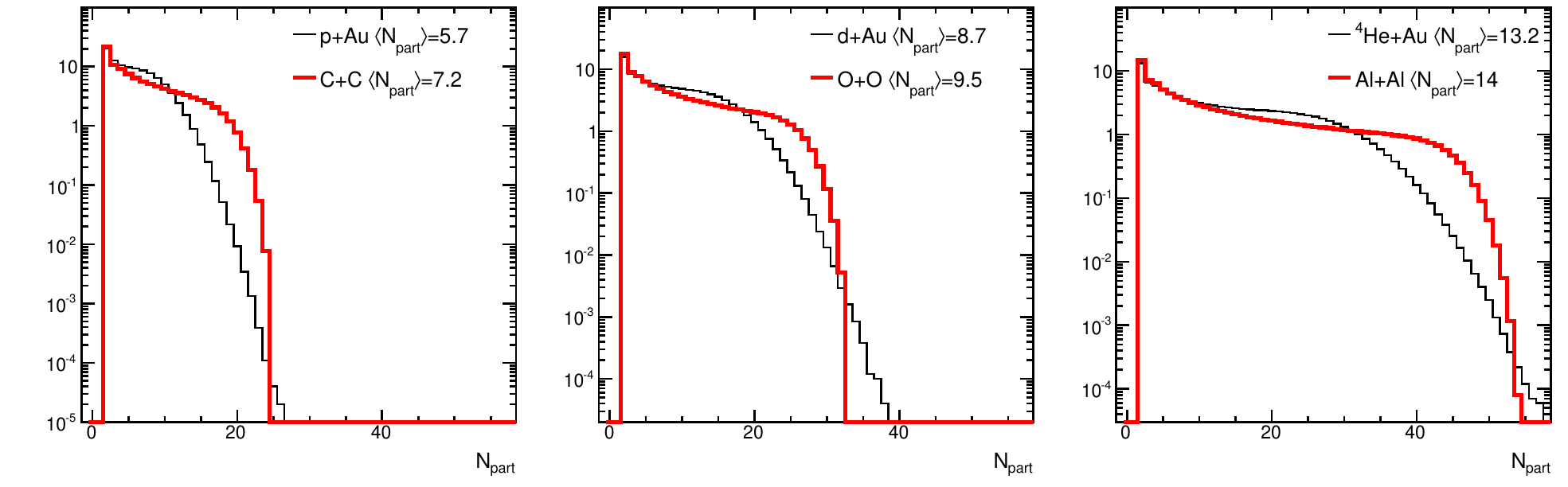}
\end{center}
\vspace*{-0.5cm}
\caption{\label{fig:1} Distributions of $\npart$ for three pairs of symmetric and asymmetric collision systems with similar $\lr{\npart}$: $p$+Au vs. C+C (left), $d$+Au vs. O+O (middle) and $^4$He+Au vs. Al+Al (right).}
\end{figure}

We consider four symmetric collision systems, $^{12}$C+$^{12}$C, $^{16}$O+$^{16}$O, $^{27}$Al+$^{27}$Al and $^{40}$Ar+$^{40}$Ar, and compare with three asymmetric systems, $p$+Au, $d$+Au and $^4$He+Au. Fig.~\ref{fig:1} shows $\npart$ distributions compared among these systems. For systems with approximately the same $\lr{\npart}$, the symmetric system has a flatter shoulder than that for the asymmetric system, which is expected to be less sensitive to experimental centrality resolution effects. 

To estimate the behavior of geometry-driven final-state collectivity in these small symmetric systems, a multi-phase transport model (AMPT)~\cite{Lin:2004en} is employed. The AMPT model has been successful in describing many features of collectivity in small- and large-system collisions at RHIC and the LHC, over a wide range of nucleus species and energies~\cite{Adare:2015cpn,Ma:2014pva,Bzdak:2014dia,Nie:2018xog}. The AMPT starts with Monte Carlo Glauber initial conditions.~\footnote{We verified with Phobos Glauber model that varying parameters such as nucleon cross-section, diffuseness parameter only has small influence on $\varepsilon_n$. The influence of alpha-clustering in $O$ was also found to be small~\cite{Lim:2018huo}} The system evolution is modeled with strings that first melt into partons, followed by elastic partonic scatterings, parton coalescence, and hadronic scatterings. The collectivity is generated mainly through elastic scatterings of partons, which leads to an emission of partons preferable along the gradient of the initial-state energy density distribution, in a manner that is similar to hydrodynamic flow. Following Refs~\cite{Ma:2014pva,Bzdak:2014dia}, we use the AMPT model v.2.25 with string-melting mode and hadronic rescatterings included. The partonic cross section of 1.5~mb is used. About 20 million AMPT events are generated for each collision system at each energy.

We compute the eccentricity vector ${\mathcal{E}}_n=\varepsilon_n {\mathrm e}^{{\textrm i}n\Psi_n}$ in each event from initial-state coordinates $(r_i,\phi_i)$ of participant nucleons as
\begin{equation}\label{eq:1}
{\mathcal{E}}_n = -\frac{\lr{r^ne^{in\phi}}}{\lr{r^n}}, n=2, 3.
\end{equation}
The phase of the eccentricity vector $\Psi_n$ is known as the participant plane (PP).

The harmonic flow coefficients $v_2$ and $v_3$ are calculated for charged particles in $0.2<\pT<3$ GeV/$c$ and $|\eta|<2.0$ using two methods. In the PP method, the anisotropy coefficients $v_n$ are calculated from the $\phi$ angles of charged hadrons relative to the $\Psi_n$, then averaged over events:
\begin{equation}\label{eq:2}
v_n\{\mathrm{PP}\} = \lr{\cos n(\phi-\Psi_n)}.
\end{equation} 
This method has the advantage of avoiding correlations from jets and resonance decays, since they are uncorrelated with the $\Psi_n$. It is used mainly to establish baseline features of $\nchb$ dependence of $v_n$ without the complication of non-flow. Since $\Psi_n$ is not experimentally accessible, and is generally different from the event plane for the final-state particles, we also calculate $v_n$ using the standard two-particle correlation (2PC) technique commonly employed in experimental measurements. In this method, harmonic coefficients are calculated from the relative azimuthal angle $\Dphi=\phi_i-\phi_j$ of pairs of charged particles as $v_{n,n}\{\mathrm{2PC}\} = \lr{\cos n(\Dphi)}$. A pseudorapidity gap of $|\Deta|>1.5$ between the pairs is required to suppress short-range correlations. To suppress non-flow correlations from back-to-back jets, a peripheral subtraction procedure, similar to that used in Ref.~\cite{Adare:2014keg}, is employed to obtain the final flow coefficients
\begin{equation}\label{eq:3}
v_n^2\{\mathrm{2PC,sub}\} = v_{n,n}\{\mathrm{2PC}\}-\frac{\lr{N_{\mathrm{ch}}}_{pp}}{\nch}v_{n,n}\{\mathrm{2PC,pp}\}\;,
\end{equation}
where the $\lr{N_{\mathrm{ch}}}_{pp}$ and $v_{n,n}\{\mathrm{2PC,pp}\}$ are the average charged particle multiplicity and harmonic coefficient from $pp$ collisions, respectively. This subtraction method was shown to work reasonably well for $\pT$-integrated correlation measurement (but underestimate the flow signal for $\pT>1-2$ GeV/$c$)~\cite{Lim:2019cys}.

The $v_n\{\mathrm{PP}\}$ and $v_n\{\mathrm{2PC,sub}\}$ are calculated as a function of centrality, which are determined based on either $\npart$ or the number of charged particles $\nch$ in the forward rapidity region $2.5<|\eta|<4.5$. In each case, the $v_n\{\mathrm{PP}\}$ and $v_n\{\mathrm{2PC,sub}\}$ are calculated in unit $\npart$ or $\nch$ bin and then averaged to obtain results in larger centrality ranges.

\begin{figure}[h!]
\begin{center}
\includegraphics[width=1\linewidth]{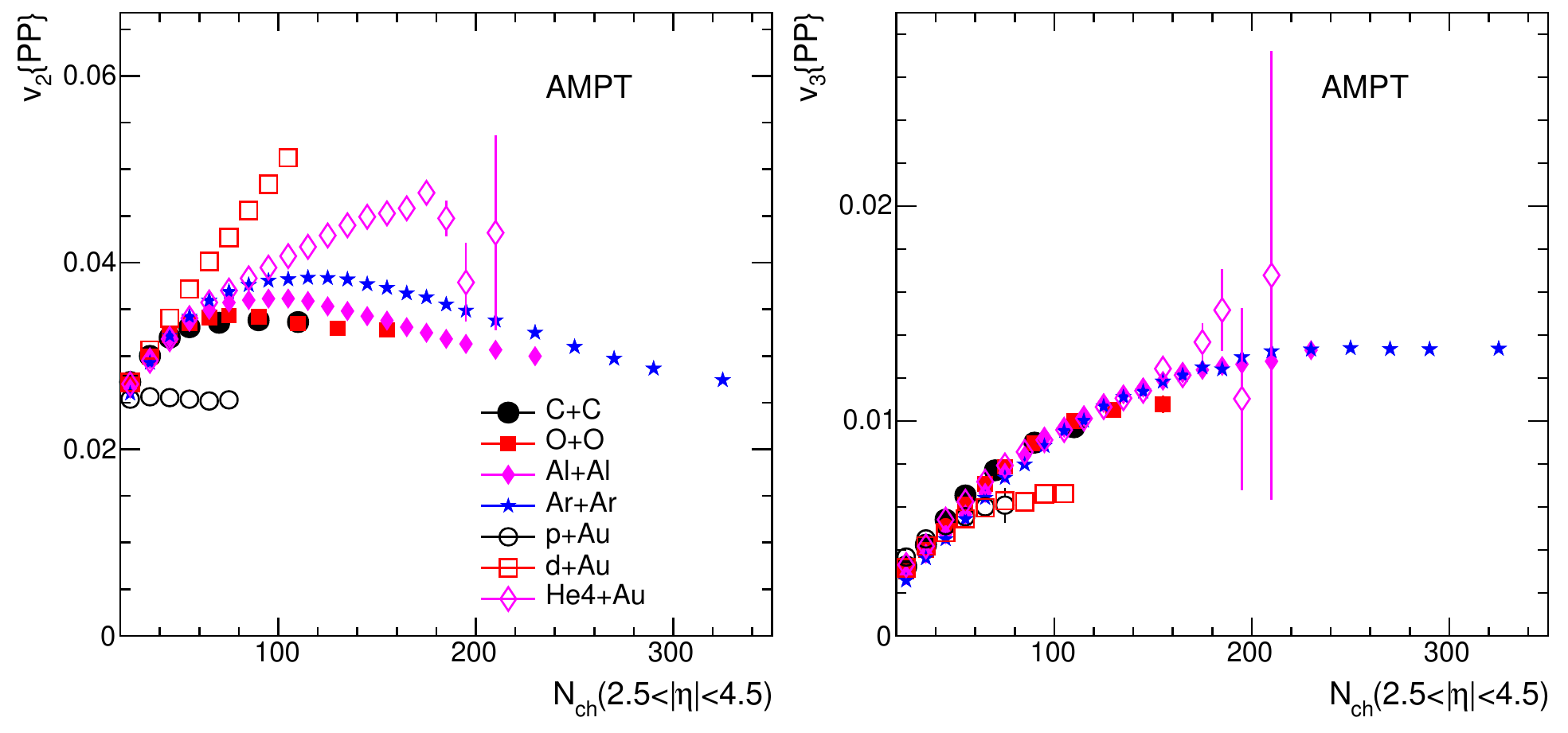}
\end{center}
\vspace*{-0.5cm}
\caption{\label{fig:2} The $v_2\{\mathrm{PP}\}$ (left) and $v_3\{\mathrm{PP}\}$ (right) as a function of $\nch$ in four symmetric and three asymmetric small collision systems.}
\end{figure}

Figure~\ref{fig:2} shows the $v_n\{\mathrm{PP}\}$ as a function of $\nch$ in four symmetric and three asymmetric small systems. For symmetric systems, the $v_2\{\mathrm{PP}\}$ values increase and then decrease with increasing $\nch$, and the peak positions in $\nch$ also increase slightly for larger systems. This behavior has been observed in larger systems~\cite{Agakishiev:2011eq,Alver:2006wh,Adare:2014bga,Adamczyk:2017ird,Adam:2019woz} and is consistent with the expectation that the $\varepsilon_2$ is driven by the average shape the overlap region~\cite{Adam:2019woz}. The $v_3\{\mathrm{PP}\}$ values for different symmetric systems tend to follow a common increasing trend as a function of $\nch$. Similar observation has been made in Cu+Cu, Au+Au and U+U collisions at RHIC~\cite{Adam:2019woz}, and in $p$+Pb and peripheral Pb+Pb collisions at the LHC~\cite{Chatrchyan:2013nka,Aaboud:2018syf}. Based on an independent source picture and a simple conformal scaling argument~\cite{Basar:2013hea}, this scaling behavior is expected since $\varepsilon_3$ is driven by random fluctuations of the positions of participating nucleons. 

Figure~\ref{fig:2} also shows that the $v_2\{\mathrm{PP}\}$ values from asymmetric systems follow different trends: the $v_2\{\mathrm{PP}\}$ in $d/^4$He+Au increase with $\nch$, while it is relatively constant in $p$+Au. The $v_3\{\mathrm{PP}\}$ values show a similar $\nch$ dependence as symmetric systems, except for $d$+Au which deviates from the common trend at large $\nch$. Therefore, in a final-state driven model, we expected a clear difference between $d/^4$He+Au and A+A for $v_2$, but relatively similar behavior for $v_3$.

\begin{figure}[h!]
\begin{center}
\includegraphics[width=1\linewidth]{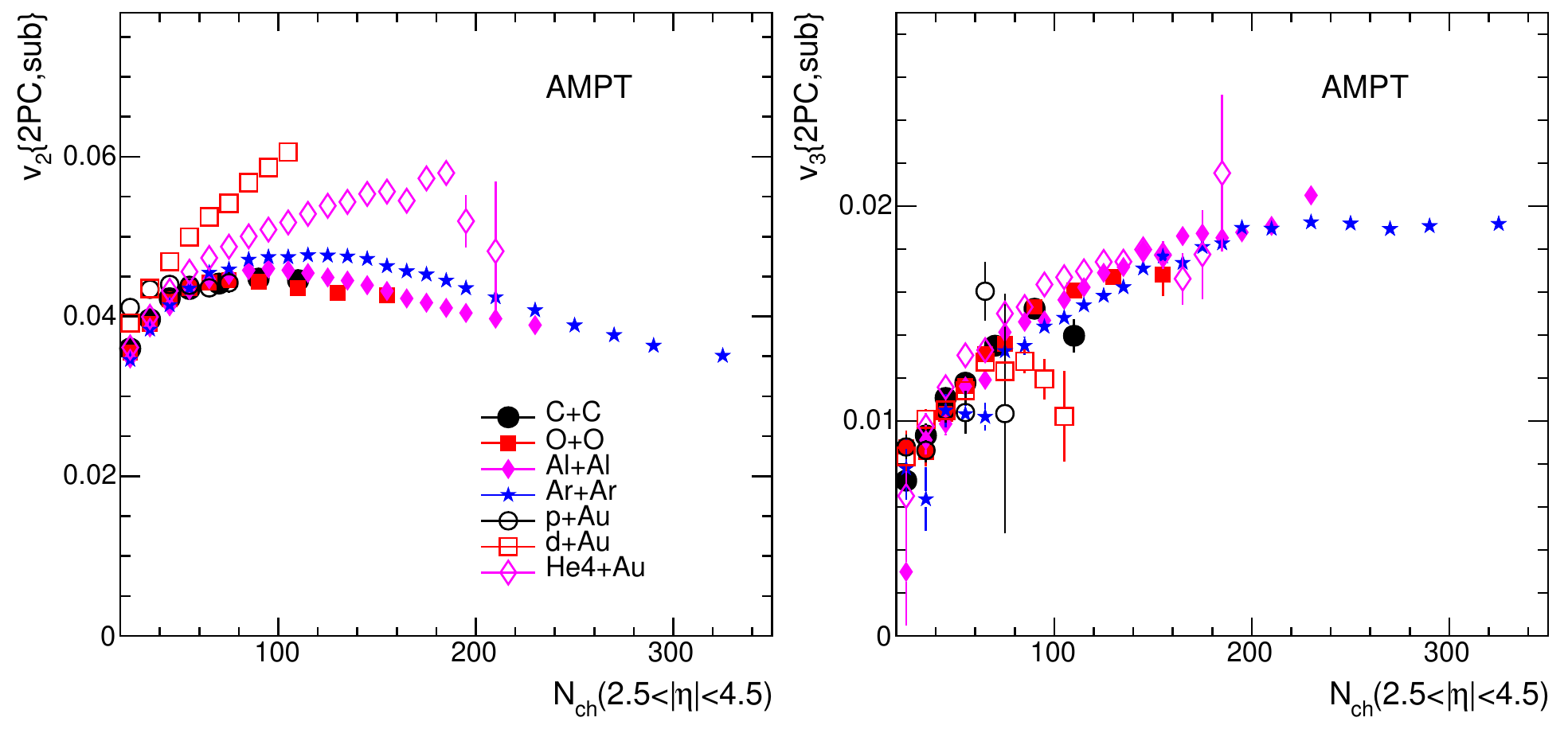}
\end{center}
\vspace*{-0.5cm}
\caption{\label{fig:3} The $v_2\{\mathrm{2PC,sub}\}$ (left) and $v_3\{\mathrm{2PC,sub}\}$ (right) as a function of $\nch$ in four symmetric and three asymmetric small collision systems.}
\end{figure}

Figure~\ref{fig:3} shows the same results for $v_n\{\mathrm{2PC,sub}\}$. The overall trends are similar to $v_n\{\mathrm{PP}\}$ in Fig.~\ref{fig:2}. The larger values of $v_n\{\mathrm{2PC,sub}\}$ are possibly due to contributions from initial momentum anisotropy that may survive to the final state in small systems, as well as possible dynamical flow fluctuations generated by final-state interactions~\cite{Aaboud:2019sma}, both of which are uncorrelated with the PP.

Since a geometry response picture is absent for pure initial momentum anisotropy models, several behaviors of $v_2$ discussed in Figs.~\ref{fig:2} and \ref{fig:3} are not naturally expected, including the $\nch$ dependence and the differences between asymmetric and symmetric systems. Therefore, measurements of centrality dependence of $v_2$ and $v_3$ and comparison with large A+A systems at similar $\nch$ can provide strong constraints on whether the observed anisotropy is dominated by initial- or final-state effects. 

In a recent yellow report for the future LHC heavy-ion physics program, the possibility for smaller A+A collisions is discussed~\cite{Citron:2018lsq}. This includes a possible $^{16}$O+$^{16}$O run at $\sqrtsnn=2.76-7$~TeV in 2022--2023, and other light-ion species such as Ar+Ar beyond 2028. The main argument for O+O run at the LHC is that it allows a better control of $\npart$, $\varepsilon_n$ and hard-scattering rate via number of nucleon-nucleon collisions $\ncoll$~\cite{Citron:2018lsq}. An O+O run at RHIC right after BES-II would provide an unprecedented and timely comparison of the same small system at very different collision energies (0.2 TeV vs. 2.76--7 TeV). This ``RHIC-LHC energy scan'' provides a unique opportunity to study systems with nearly identical nucleon geometry but very different subnucleon fluctuations and particle production mechanism with different saturation scale and mini-jet productions in the initial state. The large lever-arm in collision energy should provide new insights on the onset behavior of collectivity, jet quenching, or any other final-state effects in small systems: any model has to describe results at both energies, which naturally leads to better understandings of results at each energy. 

\begin{figure}[h!]
\begin{center}
\includegraphics[width=1\linewidth]{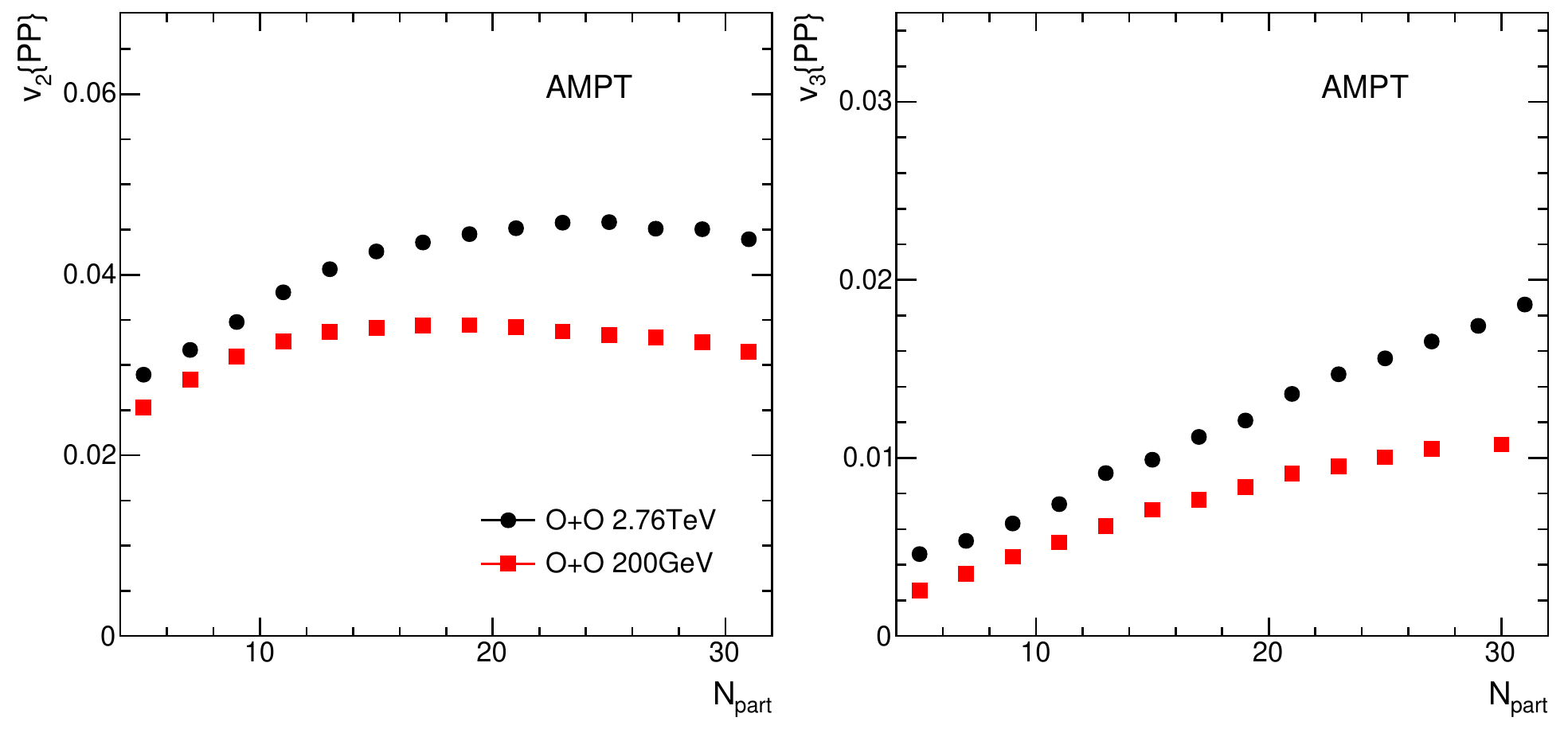}\\
\includegraphics[width=1\linewidth]{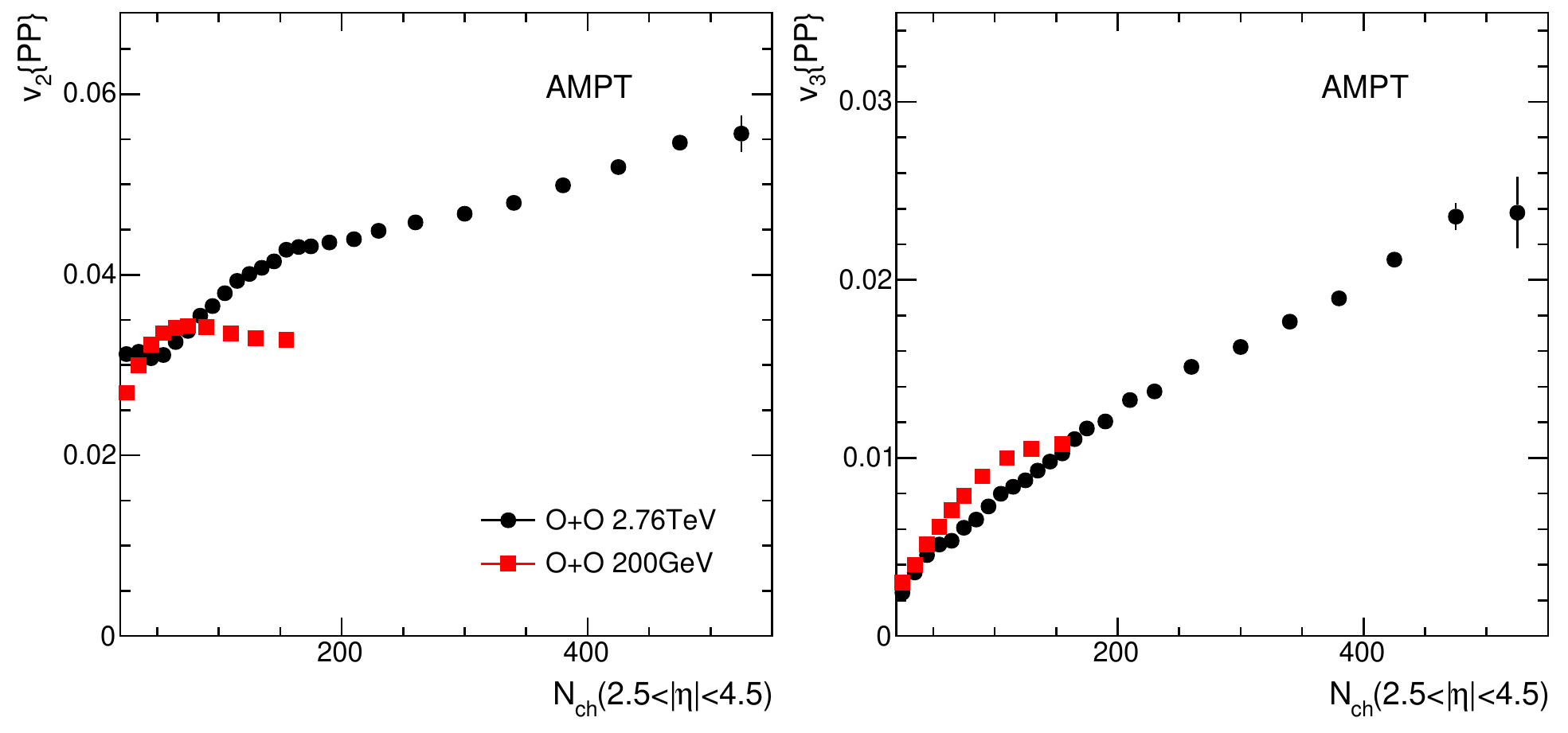}
\end{center}
\vspace*{-0.5cm}
\caption{\label{fig:4} The $v_2\{\mathrm{PP}\}$ (left) and $v_3\{\mathrm{PP}\}$ (right) as a function $\npart$ (top) or $\nch$ (bottom) for O+O collisions compared between $\sqrtsnn=0.2$ TeV and 2.76 TeV.}
\end{figure}

The top panels of Fig.~\ref{fig:4} compare the AMPT model prediction of $v_2\{\mathrm{PP}\}$ and $v_3\{\mathrm{PP}\}$ as a function of $\npart$ in O+O collisions at 0.2 and 2.76 TeV. The $v_n\{\mathrm{PP}\}$ values are larger at 2.76 TeV, but the shape of the $\npart$ dependence is rather similar between the two energies.

The bottom panels of Fig.~\ref{fig:4} show $v_n\{\mathrm{PP}\}$ as a function of $\nch$. The results for 2.76 TeV span about a factor of 2.5 larger $\nch$ range than those for 0.2 TeV, due to a larger multiplicity at a higher collision energy. More interestingly, the shape of the $\nch$ dependence of $v_2\{\mathrm{PP}\}$ is qualitatively different from its $\npart$ dependence: $v_2$ increases with $\nch$, reaching a plateau, then increases again towards higher $\nch$. The increase at larger $\nch$ resembles the behavior of $v_3\{\mathrm{PP}\}$ at large $\nch$. To offer a plausible explanation, we note that the particle production depends on fluctuations in $\npart$ and fluctuation of number of partons within each participant, and the higher-end of the $p(N_{\mathrm{ch}})$ distribution at 2.76 TeV is dominated by the fluctuation of particle production in each participant. Due to this, the $\varepsilon_n$ value is nearly constant for $\nch>300$ at 2.76 TeV (Figure 1 in supplementary materials), while it decreases continuously with $\nch$ at 0.2 TeV. Since viscous damping effects are reduced for events with the same $\varepsilon_n$ but higher multiplicity, this leads to an increases of $v_2$ with $\nch$ at 2.76 TeV in AMPT.

\begin{figure}[h!]
\begin{center}
\vspace*{-0.5cm}\hspace{-0.3cm}\includegraphics[width=0.53\linewidth]{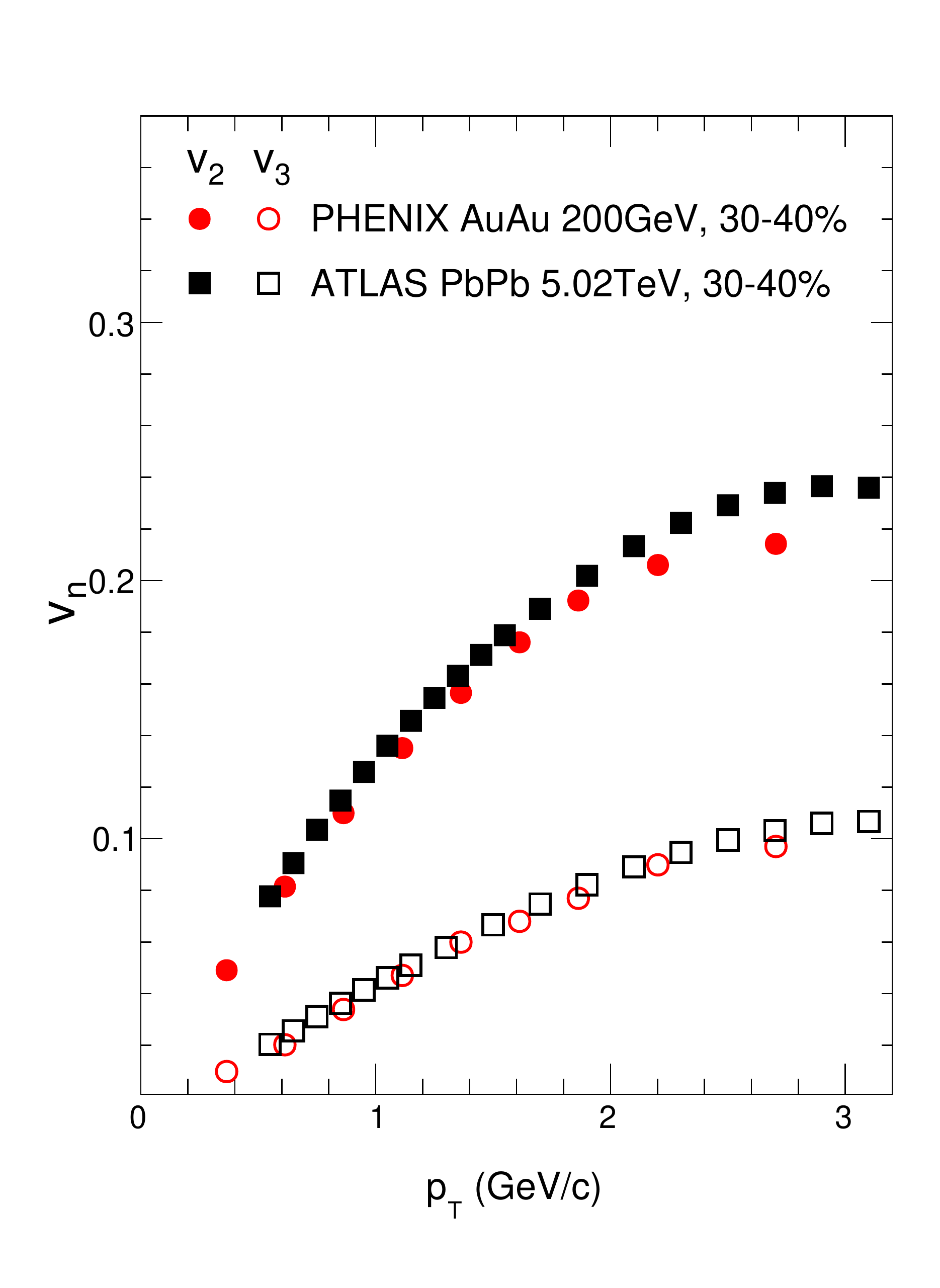}\hspace{-0.3cm}\includegraphics[width=0.53\linewidth]{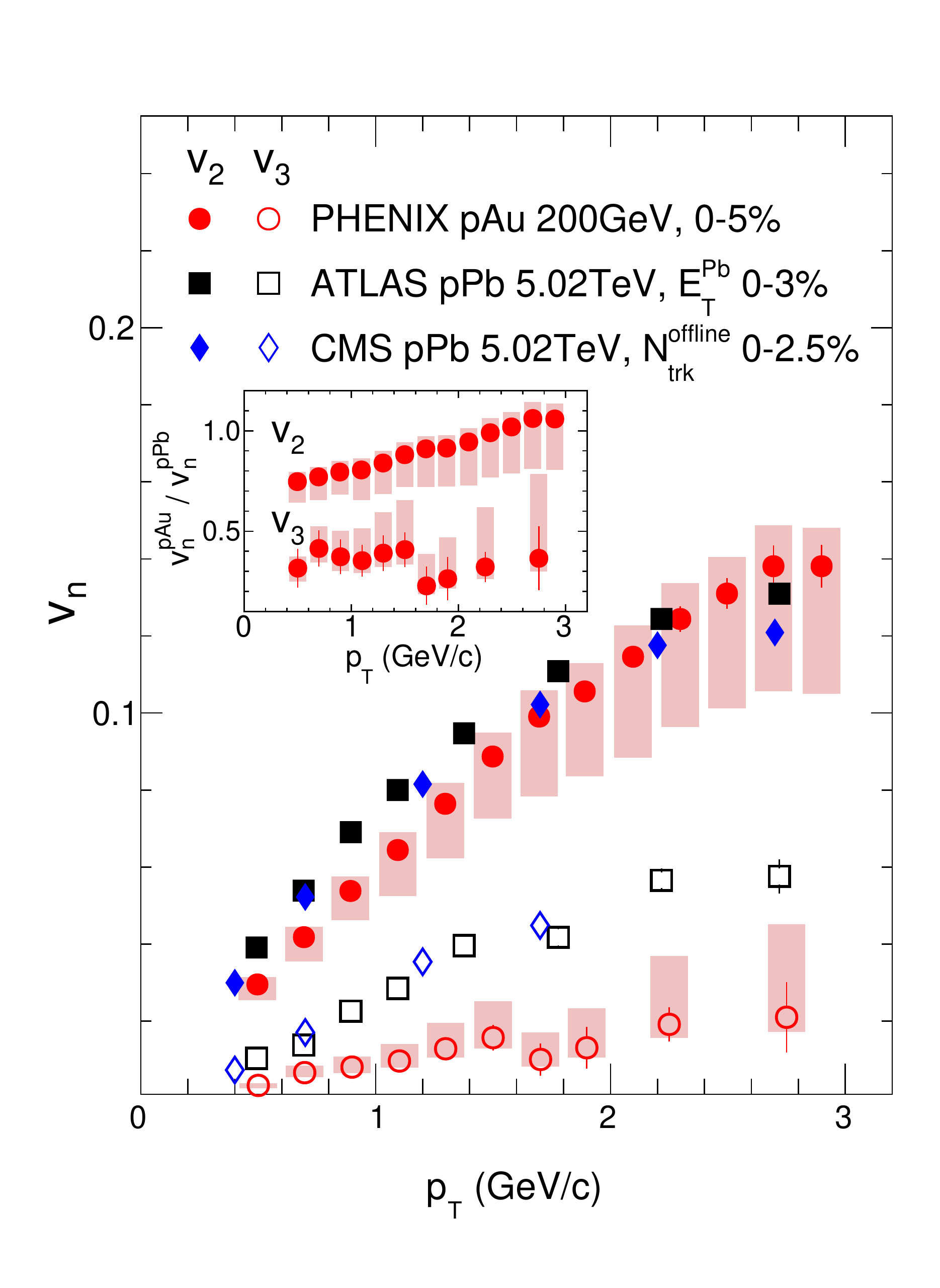}
\end{center}
\vspace*{-0.7cm}
\caption{\label{fig:5} Comparison of $v_2$ and $v_3$ between Pb+Pb and Au+Au in 30--40\% centrality (left) and between high-multiplicity $p$+Pb and $p$+Au (right). The insert panel shows the ratio of $v_n$ between $p$+Au and $p$+Pb.}
\end{figure}

To further motivate the synergy between RHIC and LHC for small system scan program, Fig.~\ref{fig:5} compares the $v_n(\pT)$ data for $n=2$,3 at two energies in large A+A system and $p$+A system. It is well-known that $v_n(\pT)$ for charged hadrons has very little $\sqrtsnn$ dependence from RHIC to LHC~\cite{Aamodt:2010pa}, as well as from 39 to 200 GeV at RHIC~\cite{Adler:2004cj,Star:2018zpt}, this is confirmed by the left panel which compares the Pb+Pb~\cite{Aaboud:2018ves} with Au+Au data~\cite{Adare:2011tg} in 30--40\% centrality. However, a comparison of $v_n(\pT)$ between $p$+Pb~\cite{Aad:2014lta} and $p$+Au~\cite{PHENIX:2018lia} central data suggests a very different story. The $v_2(\pT)$ values are more or less in agreement, but the $v_3$ at RHIC is lower by more than a factor of two and the relative difference show no apparent $\pT$ dependence. In the FSM picture, this observation suggests large change in the initial eccentricities or viscosity damping effects between the two collision energies, but the exact origin is not clear. However this observation is contested by a recent preliminary measurement from the STAR Collaboration~\cite{roy}. It would be vital to see whether the strikingly different $\sqrtsnn$ dependence for $v_2$ and $v_3$ in $p$+A collisions also persists in small A+A systems such as O+O collisions between RHIC and LHC.

The large gap between $pp$ and Cu+Cu is one of the last unexplored frontiers at RHIC~\cite{wolfram} and now is the best time to fill it. Since the last RHIC $p$/$d$/He+Au scan, the STAR experiment has completed several detector upgrades that extend $\pT$ and particle identification to $|\eta| < 1.5$, and provide centrality and event plane determination in $2<|\eta|<5$; An ongoing forward upgrade to instrument $2.5 <\eta< 4$ region with tracking detector and calorimeter is expected to complete in 2021~\cite{STAR}. A one-week 200 GeV O+O run was recently proposed by the STAR Collaboration for 2020 or 2021~\cite{starbup}, which is expected to provide 400 million minimum bias events and 200 million 0--5\% central events. This dataset would enable detailed measurements of multi-particle correlations and rare particles such as $\phi$ meson with decent precision. A forward upgrade has also been planned for the sPHENIX experiment~\cite{Adare:2015kwa}. The extended detector capability should allow a full exploration of collectivity using all the observable and methods developed for large systems at RHIC/LHC. We will have much better control of the non-flow systematics, understanding the multi-particle nature of the collectivity and the longitudinal correlations to constrain the full 3D initial condition. 

In summary, we propose a scan of small A+A systems at RHIC top energy $\sqrtsnn=200$ GeV to understand the timescale for the emergence of collectivity and early thermalization mechanisms in nucleus-nucleus collisions. Comparing to asymmetric systems such as $p$/$d$/He+Au with similar $\npart$, the symmetric systems have different initial geometry fluctuations and less bias on the centrality selection. A scan of both symmetric and asymmetric systems provide an opportunity to disentangle contributions to collectivity from initial momentum anisotropy, pre-equilibrium and late-time hydrodynamics, as well as to study the onset of other final-state effects such as jet quenching. An O+O run at RHIC to match an already planned LHC O+O run around 2021--2022 will for the first time probe the nature of collectivity with the same nucleon geometry and size but very different subnucleon fluctuations and space-time evolution due to the $\times 13-30$ difference in the collision energy.

We appreciate valuable discussions with Fuqiang Wang, Li Yi, Prithwish Tribedy, Roy Lacey, Aihong Tang and ShinIchi Esumi. We acknowledges the support from NSF grant number PHY-1613294 and PHY-1913138 (JJ), DOE grant number DE-FG02-87ER40331.A008 (SH,ZC) and DE-SC0012185 (WL).

\bibliography{sysscan}{}
\bibliographystyle{apsrev4-1}
\end{document}